# Black Magic in Gray Titania: Noble-Metal-Free Photocatalytic H2 Evolution from Hydrogenated Anatase

*Ning Liu[1], Xuemei Zhou[1], Nhat Truong Nguyen[1], Kristina Peters[2], Christopher Schneider[1], Detlef Freitag[3], Dina Fattakhova-Rohlfing[2], Patrik Schmuki[1,4]\**

[1]*Department of Materials Science WW-4, LKO, University of Erlangen-Nuremberg, Martensstrasse 7, 91058 Erlangen, Germany;*

[2]*Department of Chemistry and Center for NanoScience (CeNS), Ludwig-Maximilians-Universität München, Butenandtstr. 5-11 (Haus E), 81377 Munich, Germany;*

[3]*High Pressure Laboratory, Chair of Separation Science and Technology, University of Erlangen-Nuernberg, Haberstrasse 11, 91058 Erlangen, Germany;*

[4]*Department of Chemistry, King Abdulaziz University, Jeddah, Saudi Arabia.*

*\*Corresponding author. Tel.: +49 91318517575, fax: +49 9131 852 7582*

*Email:* [schmuki@ww.uni-erlangen.de](mailto:schmuki@ww.uni-erlangen.de)

'Black' TiO$_2$ has gained increasing interest because of its outstanding properties and promising applications in a wide range of fields. Among the outstanding features of the material is that certain synthesis processes lead to the formation of an intrinsic co-catalytic center and thus enable noble-metal free photocatalytic H$_2$ generation. In this work, we report 'grey TiO$_2$' by an appropriate hydrogenation treatment exhibits excellent photocatalytic hydrogen. In this case, by the employment of thermally stable and high-surface-area TiO$_2$ nanoparticles as well as mesoporous particles as the hydrogenation precursor, the appropriate extent of reduction of TiO$_2$ (coloration) and the formation of Ti$^{3+}$ is the key for the efficient noble-metal-free photocatalytic H$_2$ generation. The EPR results reveal that 'grey TiO$_2$' shows stronger Ti$^{3+}$ feature at g $\approx$ 1.93 than 'black TiO$_2$' contributing to the intrinsic catalytic center for H$_2$ evolution.

Ever since Fujishima and Honda used TiO$_2$ as a photoanode in a photoelectrochemical cell (PEC) for hydrogen production [1], TiO$_2$ remains the most commonly used wide-band gap semiconductor photoanode for water splitting. The main reasons are that the material provides suitable band-edge positions relative to the water red-ox potentials, low cost, and a very high photocorrosion resistance [2-4].

However, the main weaknesses of TiO$_2$ are its large band gap (3.0-3.2 eV) and a very sluggish transfer kinetics of the charge carriers to the surrounding media. To achieve higher light conversion a broad range of approaches has been explored to narrow the TiO$_2$ optical band gap – such band gap engineering approaches have been based on the introduction of suitable energy states into the TiO$_2$ gap using doping elements (such as N, C, Cr or W) [4, 5]. To improve the slow electron transfer kinetics of TiO$_2$ to the adsorbed hydrogen species, the common remedy for accelerating the $H_2$ generation is decorating the $TiO_2$ particles with co-catalyst particles, typically Pt, Au or Pd etc. that catalyze charge separation and hydrogen evolution [6, 7].

In the context of achieving a broad visible light absorption, a remarkable finding was reported in 2011when Chen and Mao introduced so-called 'black titania' [8]. The material showed in optical absorption measurements a band gap of 1.54 eV, and after decoration with Pt as co-catalyst, a remarkably high water splitting performance [8]. This 'black' modification of $TiO_2$ was produced by exposing anatase nanoparticles to a high pressure/high temperature treatment in $H_2$ (HPH). Recently, we used this HPH treatment on $TiO_2$ nanotube arrays and commercial anatase particles and found another compelling feature of this material, which is that these modified structures provide a high and stable open-circuit photocatalytic hydrogen production rate without the presence of any noble-metal co-catalyst [9, 10]. Based on EPR and PL measurements we ascribed the intrinsic co-catalytic activation of anatase $TiO_2$ to specific defect centers formed during hydrogenation.

Despite these achievements, it still remains unclear whether the more coloration of $TiO_2$ (larger solar absorption) means more efficient for the photocatalytic activities. Moreover, the performance of 'black $TiO_2$' was reported different depending on different hydrogenation methods [16-19]. Hence a clear assessment of the key factors in the treatment to the enhanced photoactivities must be explored.

In the present work, anatase nanoparticles with different hydrogenation treatment (different temperature, flow and high pressure $H_2$ atmosphere) show significantly different photocatalytic activity for noble-metal-free $H_2$ generation. It shows that an appropriate hydrogenation treatment can lead to 'grey $TiO_2$' with more activation for noble-metal-free $H_2$ generation than 'black $TiO_2$', even the 'black $TiO_2$' shows much stronger solar absorption. Moreover EPR result for 'grey TiO2' shows stronger Ti3+ peak ascribed the intrinsic co-catalytic activation than 'black TiO2'.

Hydrogenated $TiO_2$ nanoparticles were obtained by annealing pristine anatase $TiO_2$ nanoparticles (25 nm, Sigma-Aldrich) under $H_2$ flow and 20 bar at different temperature ranging 400-700 °C for 1 h using the flow furnace and the autoclave for high pressure (for details, see the Supporting Information).

To assess the ability of the different hydrogenated particles to produce photocatalytic $H_2$ without using any co-catalyst, we irradiated the particles under open circuit voltage (OCV) conditions in an aqueous methanol solution (50 vol %) under AM 1.5 (100 mW/cm2) solar simulator conditions (Fig. 1a). It can be seen that the hydrogenated particle show a significant photocatalytic $H_2$ evolution activity. As reference for a classic activation treatment for $TiO_2$, we include results for high pressure annealing in $H_2$ (HPH). The best result from the particles treated at 500 °C in flow $H_2$ is as efficient as the classic HPH result for noble-metal-free $H_2$ evolution. Nevertheless, the particles treated at higher temperature decay the efficiency gradually. The non-treated anatase does not produce any considerable amount of $H_2$.

The different hydrogenation treatment leads to different coloration of the particles and considerable absorption in the visible spectral range (see Fig. 1b). It can be seen for the photographs in Fig.1b that the sample treated at 400 °C still maintains white color as the pristine $TiO_2$. The particles treated at 500 °C can be observed slightly grey color while the color of sample changes to more black after 600 °C hydrogenation and 'black' (black blue) $TiO_2$ was obtained at 700 °C. Fig. 1b shows the absorption in the visible light region gradually increased with the hydrogenation temperature in UV-vis absorption spectra. Absorption can be observed up to 800 nm. The particles treated at 700 °C provide the strongest absorption. The bandgap energy of the hydrogenated $TiO_2$ samples becomes lower with the increasing heating temperature, which is obtained on the basis of the Kubelka-Munk function in Fig. 1c.

These findings indicate that the magnitude in optical absorption (in the visible or UV range) does not correlate with the observed open-circuit photocatalytic $H_2$ production activity in Fig.

1a. Clearly, 'grey $TiO_2$' under OCV condition shows a significant more amount of photocatalytic H2 production than 'black $TiO_2$'. Even more, the 'black' particles treated at 700 °C shows the lowest efficiency for noble-metal-free water splitting with the narrowest band-gap in Fig. 1c.

The X-ray diffraction (XRD) patterns of all the samples before and after hydrogenation treatments in Fig. 2a. The samples treated at the elevated temperature hydrogenation (500 and 600 ˚C) show pure anatase phase as the pristine anatase. The particles with the 700 ˚C hydrogenation show a major anatase phase with a small amount of rutile phase. Notably, the peak intensity of anantase significantly decreased after hydrogenation, which is in line with our finding for HPH samples with Rietveld refinement analysis [10]. The average crystallite size (using the Scherrer equation calculated) of the hydrogenated is appx. 20 nm, which is smaller than that of pristine powder (30 nm).

Further Raman spectra were acquired to investigate the structural changes of $TiO_2$ after hydrogenation in Fig. 2b. After hydrogenation, Raman bands at the main effect is a mild blue shift, evident e.g. at the main Eg peak, that is at 134 $cm^{-1}$ for the pristine powders to 144 $cm^{-1}$ for the hydrogenated powders. These observations are in line with models that indicate phonon confinement by a decrease of the effective particle size. Such a reduction of crystallite size could be due to amorphization of the original lattice induced by the $H_2$, if for instance as observed in literature [8, 12] the amorphous shell around the particles would increase. However, our TEM of the hydrogenated particles at different temperature did not show a significant increase of the thickness of these amorphous shells (Fig. 2c-g). TEM images for the sample at 700 ˚C hydrogenation shows an aggregation in the nanoparticle morphology. Moreover, from XPS (in Fig. 2d) no significant variation in the composition, such as $Ti^{3+}$, of the samples before and after hydrogenation could be observed.

In order to gain additional information on the electronic nature of the defect structure, electron paramagnetic resonance (EPR) measurements were carried out. From EPR spectra taken for different samples at 80 K in dark, a different defect signature becomes apparent (Fig. 3). All the samples, except the sample treated at 700 ˚C show the peaks with a g-factor of 2.003 which in the literature [13] are typically assigned to oxygen vacancies (OV). The 'grey' samples show the presence of a EPR signal of g-value of 1.91 as paramagnetic defect in the dark conditions. The feature has been ascribed to a $Ti^{3+}$ species existing in the subsurface of $TiO_2$ [9, 10], contributing to the intrinsic catalytic center for $H_2$ evolution. Moreover, the sample treated at 500 ˚C shows much stronger signals at g-value of 1.91 and 2.003 than that for the sample treated at 600 ˚C which is in line with the co-catalytic activity for $H_2$ evolution. Interestingly, 'black $TiO_2$' shows a very high intensity peak at 1.957, which reveals the amount of $Ti^{3+}$ species in the bulk [14, 15]. Combined with water splitting results, the high concentration of $Ti^{3+}$ species in 'black $TiO_2$' leads to the severe bulk defects, which act as charge recombination centers.

Additionally, the photocatalytic performance of $TiO_2$ for water splitting application depends, as any surface controlled reaction directly on the number of accessible active surface sites. The 3D-titania mesoporous structures feature high accessible surface areas in combination with the interconnected network of a bulk titania phase [11, 12]. Thus we also investigate the noble-metal-free water splitting behavior of mesoporous $TiO_2$ with large surface area (approx. 130 m²/g, compared with the surface area of $TiO_2$ nanoparticles 50 m²/g, SI), confined porous structure and high surface to volume ratio by different hydrogenation treatments. Fig. 4a and b show the TEM images of mesoporous anatase with dense pore walls, being composed of sintered, interconnected nanoparticles of particle size of approx. 8 nm. After typical hydrogenation leads to 'grey TiO2', the size of the crystalline domains increased from 8 nm to 13.6 in Fig. 4c and d. Additionally, for both samples, HRTEM cannot obviously reveal the presence of amorphous regions. Fig. 4e shows the amount of $H_2$ produced under open-circuit conditions from the 'grey TiO$_2$' and 'black TiO$_2$' (as shown in photograph in Fig. 4e) of mesoporous structure comparing with that of 'grey TiO$_2$' of nanoparticles. In the mesoporous case, the 'grey TiO$_2$' shows the higher H$_2$ evolution than 'black TiO$_2$' as expected. Moreover, the mesoporous 'grey TiO$_2$' shows about 2 times higher H$_2$ evolution rate than 'grey' nanoparticles because of the larger surface area.

In summary, in the present work, we investigated 'grey TiO$_2$' and 'black TiO$_2$' resulting from different hydrogenation of nanostructure and mesoporous structure and their efficiency towards photocatalytic H$_2$ evolution. The results show that photocatalytic efficiency for noble-metal-free H$_2$ evolution is not related to the coloration of TiO$_2$ by different hydrogenation reduction. The 'grey TiO$_2$' shows stronger Ti$^{3+}$ feature at g $\approx$ 1.93 contributing to the intrinsic catalytic center in hydrogenated materials for H$_2$ evolution. The 'black TiO$_2$' treated at higher temperature shows highly concentrated bulk defects at g $\approx$ 1.96 of EPR results, which act as the recombination centers inhibiting the photoactivities. The temperature for hydrogenation is the determining factor for the photocatalytic activity for noble-metal-free water splitting both in flow and high pressure cases.

**Figure captions**:

Fig. 1 (a) Photocatalytic hydrogen evolution rate from $TiO_2$ nanoparticles with different hydrogenation treatments under AM 1.5 (100 mW/cm$^2$) illumination; (b) Integrated light reflectance results of $TiO_2$ with different hydrogenation treatments with corresponding photographs of the powders; (c) The plot of the transformed Kubella-Munk function vs. the gap energy of the $TiO_2$ samples.

Fig. 2 (a) The X-ray diffraction patterns and (b) Raman spectra of $TiO_2$ before and after hydrogenation; (c) TEM images and (d) XPS of $TiO_2$ nanoparticles before and after hydrogenation .

Fig. 3 EPR spectra for TiO$_2$ nanoparticles with different hydrogenation treatments at 80 K in dark.

Fig. 4 TEM images of mesoporous TiO$_2$ (a+b) and hydrogenated TiO$_2$ (c+d) showing the crystalline pore walls assembled from nanocrystals. The inset in (a) and (d) shows the corresponding selected area electron diffraction (SAED) pattern, which is in a good agreement with the TiO$_2$ anatase structure; (e) Photocatalytic hydrogen evolution rate from TiO$_2$ mesoporous materials with different treatments compared with 'grey TiO2' under AM 1.5 (100 mW/cm$^2$) illumination.

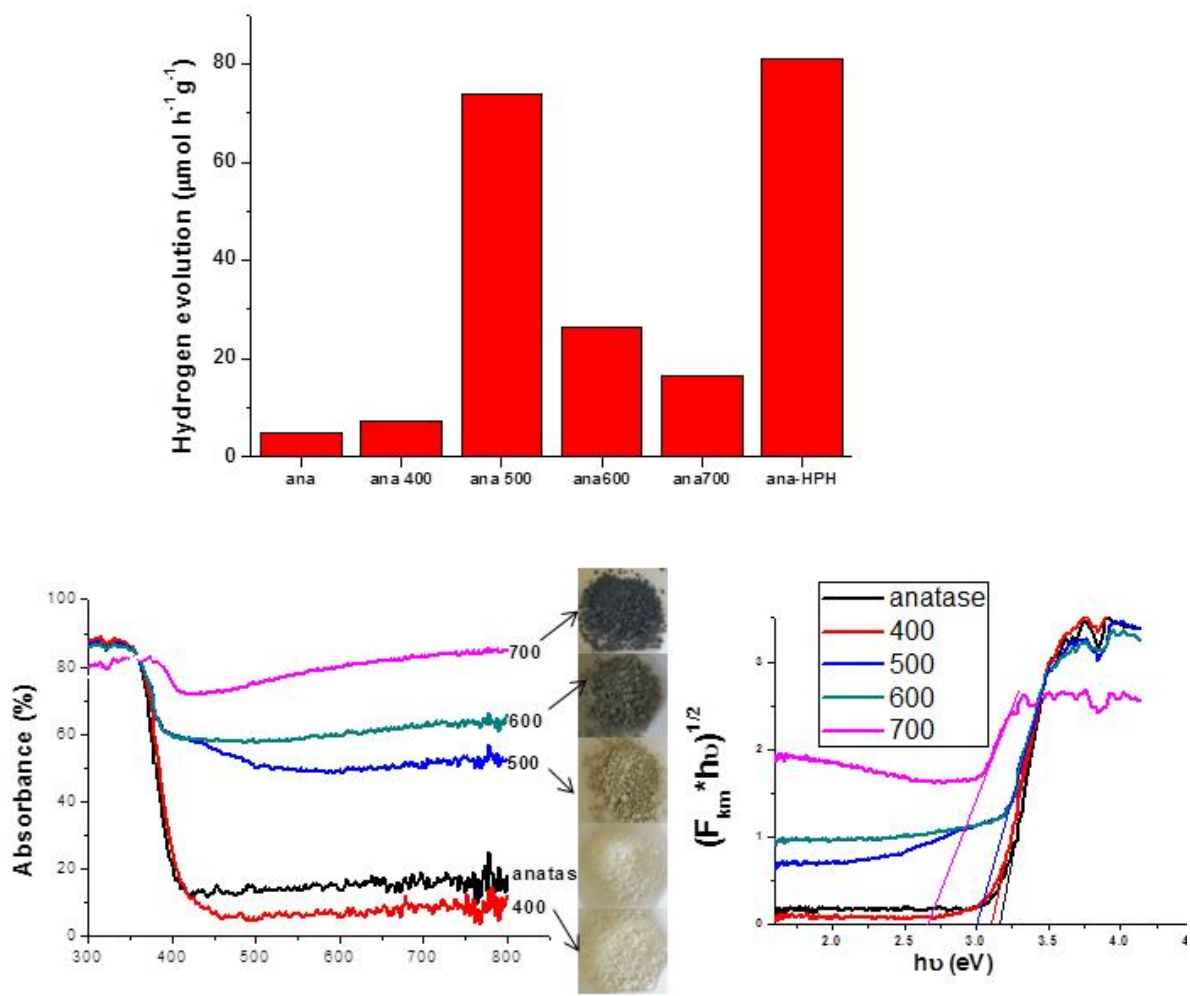

Fig. 1

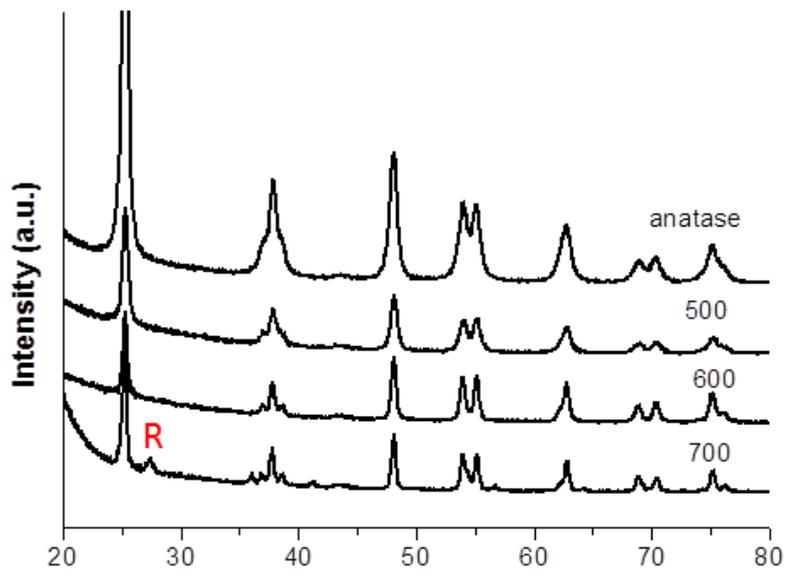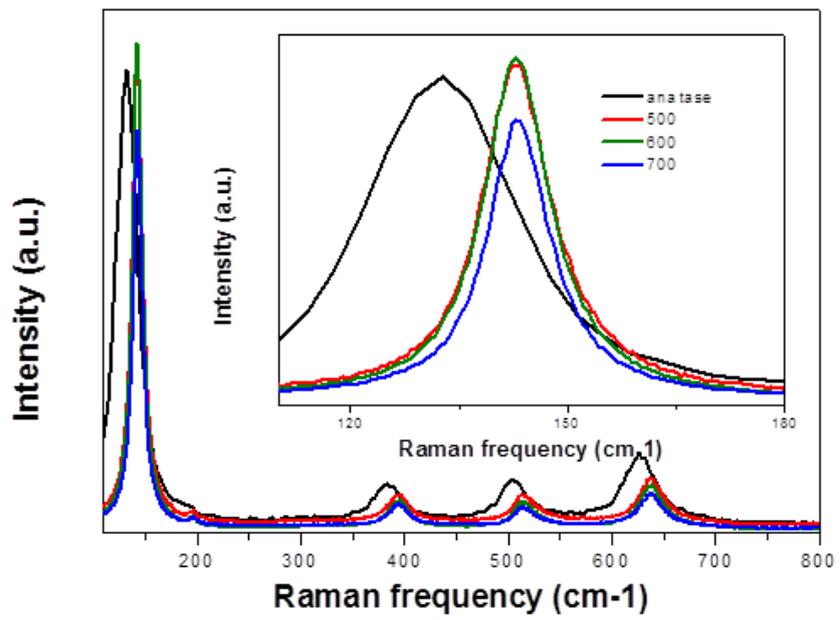

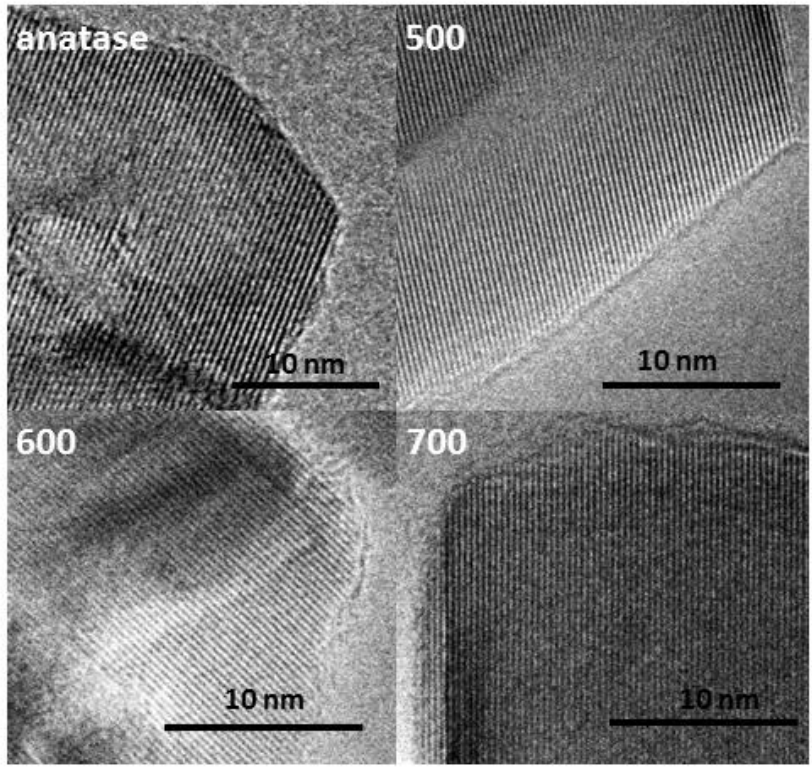

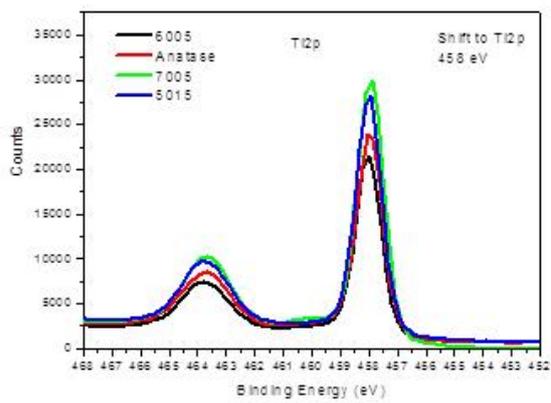
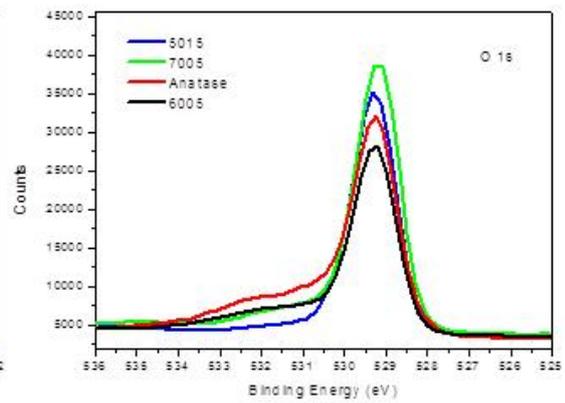

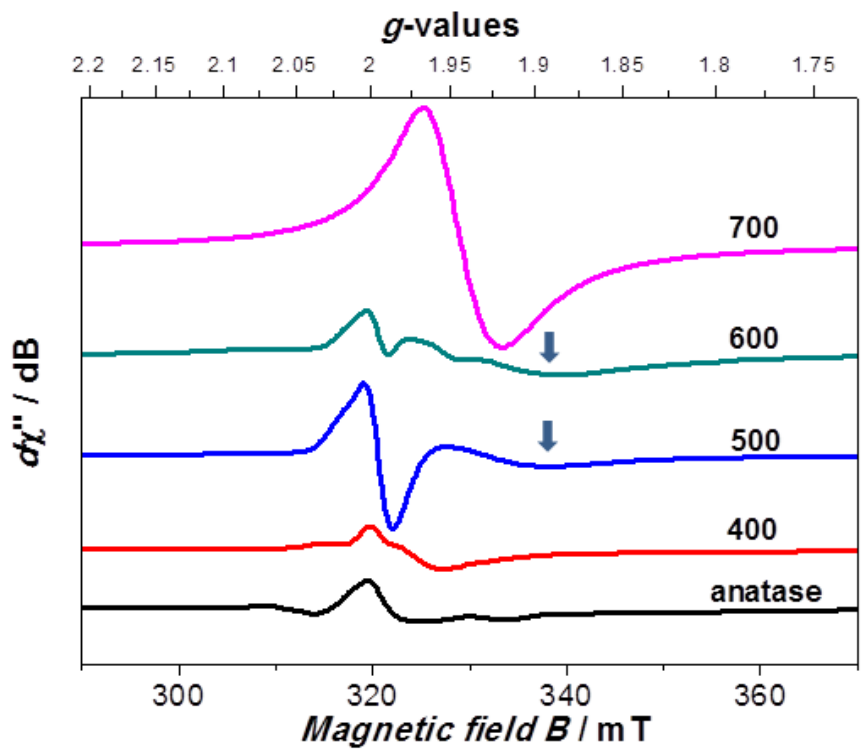

Fig. 3

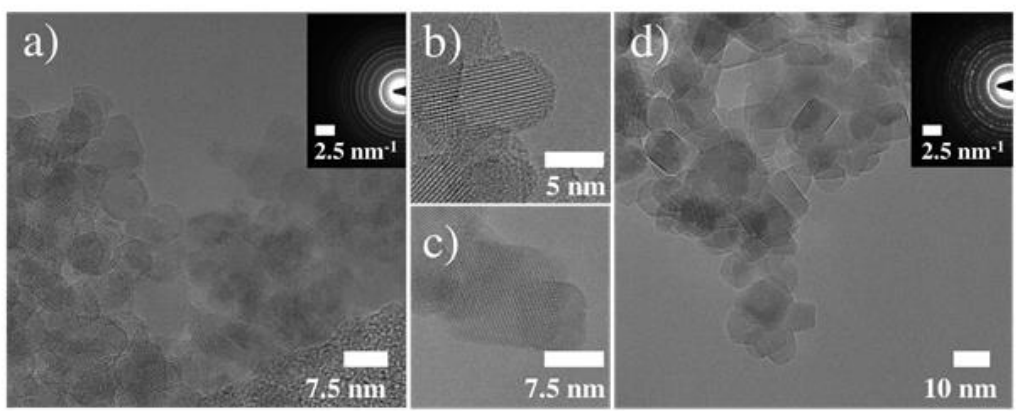
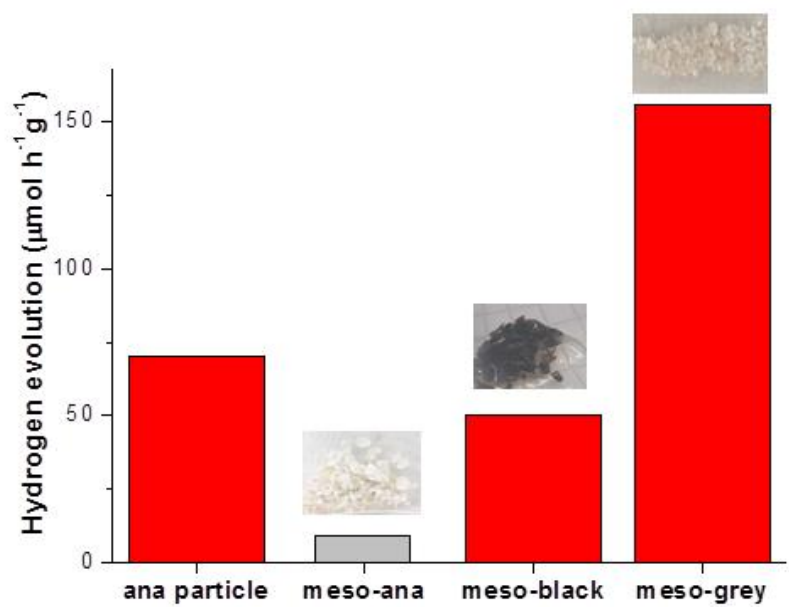

Fig. 4